
\magnification=\magstephalf
%
\newbox\SlashedBox
\def\slashed#1{\setbox\SlashedBox=\hbox{#1}
\hbox to 0pt{\hbox to 1\wd\SlashedBox{\hfil/\hfil}\hss}#1}
\def\hboxtosizeof#1#2{\setbox\SlashedBox=\hbox{#1}
\hbox to 1\wd\SlashedBox{#2}}

\def\mathslashed#1{\setbox\SlashedBox=\hbox{$#1$}
\hbox to 0pt{\hbox to 1\wd\SlashedBox{\hfil/\hfil}\hss}#1}

\def\ifsmall{\iffalse}  
\def\titlepagefont{}  

\def\DefineTeXgraphics{%
\special{ps::[global] /TeXgraphics { } def}}  

\def\today{\ifcase\month\or January\or February\or March\or April\or May
\or June\or July\or August\or September\or October\or November\or
December\fi\space\number\day, \number\year}
\def\eatPrefix19{}
\def\Year{\expandafter\eatPrefix\the\year}
\newcount\hours \newcount\minutes
\def\monthname{\ifcase\month\or
January\or February\or March\or April\or May\or June\or July\or
August\or September\or October\or November\or December\fi}
\def\shortmonthname{\ifcase\month\or
Jan\or Feb\or Mar\or Apr\or May\or Jun\or Jul\or
Aug\or Sep\or Oct\or Nov\or Dec\fi}

\def\TimeStamp{\hours\the\time\divide\hours by60%
\minutes -\the\time\divide\minutes by60\multiply\minutes by60%
\advance\minutes by\the\time%
${\rm \shortmonthname}\cdot\if\day<10{}0\fi\the\day\cdot\the\year%
\qquad\the\hours:\if\minutes<10{}0\fi\the\minutes$}




\def\Title#1{%
\vskip 1in{\titlefont\centerline{#1}}\vskip .5in}

\def\Date#1{\leftline{#1}\tenrm\supereject%
\global\hsize=\hsbody\global\hoffset=\hbodyoffset%
\footline={\hss\tenrm\folio\hss}}

\newif\ifdraftmode
\newif\ifleftlabels  

\def\nolabels{\def\wrlabeL##1{}\def\eqlabeL##1{}\def\reflabeL##1{}}
\def\writelabels{\def\wrlabeL##1{\leavevmode\vadjust{\rlap{\smash%
{\line{{\escapechar=` \hfill\rlap{\sevenrm\hskip.03in\string##1}}}}}}}%
\def\eqlabeL##1{{\escapechar-1\rlap{\sevenrm\hskip.05in\string##1}}}%
\def\reflabeL##1{\noexpand\rlap{\noexpand\sevenrm[\string##1]}}}
\def\writeleftlabels{\def\wrlabeL##1{\leavevmode\vadjust{\rlap{\smash%
{\line{{\escapechar=` \hfill\rlap{\sevenrm\hskip.03in\string##1}}}}}}}%
\def\eqlabeL##1{{\escapechar-1%
\rlap{\sixrm\hskip.05in\string##1}%
\llap{\sevenrm\string##1\hskip.03in\hbox to \hsize{}}}}%
\def\reflabeL##1{\noexpand\rlap{\noexpand\sevenrm[\string##1]}}}
\nolabels

\newdimen\fullhsize
\newdimen\hstitle
\hstitle=\hsize 
\newdimen\hsbody
\hsbody=\hsize 
\newdimen\hbodyoffset
\hbodyoffset=\hoffset 
\newbox\leftpage
\def\abstract#1{#1}
\def\rotated{\special{ps: landscape}
\magnification=1000  
\baselineskip=14pt
\global\hstitle=9truein\global\hsbody=4.75truein
\global\vsize=7truein\global\voffset=-.31truein
\global\hoffset=-0.54in\global\hbodyoffset=-.54truein
\global\fullhsize=10truein
\def\DefineTeXgraphics{%
\special{ps::[global]
/TeXgraphics {currentpoint translate 0.7 0.7 scale
              -80 0.72 mul -1000 0.72 mul translate} def}}
\let\lr=L
\def\ifsmall{\iftrue}
\def\titlepagefont{\twelvepoint}
\trueseventeenpoint
\def\almostshipout##1{\if L\lr \count1=1
      \global\setbox\leftpage=##1 \global\let\lr=R
   \else \count1=2
      \shipout\vbox{\hbox to\fullhsize{\box\leftpage\hfil##1}}
      \global\let\lr=L\fi}

\output={\ifnum\count0=1 
 \shipout\vbox{\hbox to \fullhsize{\hfill\pagebody\hfill}}\advancepageno
 \else
 \almostshipout{\leftline{\vbox{\pagebody\makefootline}}}\advancepageno
 \fi}

\def\abstract##1{{\leftskip=1.5in\rightskip=1.5in ##1\par}} }

\def\linemessage#1{\immediate\write16{#1}}

\global\newcount\secno \global\secno=0
\global\newcount\appno \global\appno=0
\global\newcount\meqno \global\meqno=1
\global\newcount\subsecno \global\subsecno=0
\global\newcount\figno \global\figno=0

\newif\ifAnyCounterChanged
\let\terminator=\relax
\def\normalize#1{\ifx#1\terminator\let\next=\relax\else%
\if#1i\aftergroup i\else\if#1v\aftergroup v\else\if#1x\aftergroup x%
\else\if#1l\aftergroup l\else\if#1c\aftergroup c\else%
\if#1m\aftergroup m\else%
\if#1I\aftergroup I\else\if#1V\aftergroup V\else\if#1X\aftergroup X%
\else\if#1L\aftergroup L\else\if#1C\aftergroup C\else%
\if#1M\aftergroup M\else\aftergroup#1\fi\fi\fi\fi\fi\fi\fi\fi\fi\fi\fi\fi%
\let\next=\normalize\fi%
\next}
\def\makeNormal#1#2{\def\doNormalDef{\edef#1}\begingroup%
\aftergroup\doNormalDef\aftergroup{\normalize#2\terminator\aftergroup}%
\endgroup}

\def\warnIfChanged#1#2{%
\ifundef#1
\else\begingroup%
\edef\oldDefinitionOfCounter{#1}\edef\newDefinitionOfCounter{#2}%
\ifx\oldDefinitionOfCounter\newDefinitionOfCounter%
\else%
\linemessage{Warning: definition of \noexpand#1 has changed.}%
\global\AnyCounterChangedtrue\fi\endgroup\fi}

\def\Section#1{\global\advance\secno by1\relax\global\meqno=1%
\global\subsecno=0%
\bigbreak\bigskip
\centerline{\twelvepoint \bf %
\the\secno. #1}%
\par\nobreak\medskip\nobreak}
\def\tagsection#1{%
\warnIfChanged#1{\the\secno}%
\xdef#1{\the\secno}%
\ifWritingAuxFile\immediate\write\auxfile{\noexpand\xdef\noexpand#1{#1}}\fi%
}
\def\section{\Section}
\def\Subsection#1{\global\advance\subsecno by1\relax\medskip %
\leftline{\bf\the\secno.\the\subsecno\ #1}%
\par\nobreak\smallskip\nobreak}
\def\tagsubsection#1{%
\warnIfChanged#1{\the\secno.\the\subsecno}%
\xdef#1{\the\secno.\the\subsecno}%
\ifWritingAuxFile\immediate\write\auxfile{\noexpand\xdef\noexpand#1{#1}}\fi%
}

\def\subsection{\Subsection}

\def\romappno{\uppercase\expandafter{\romannumeral\appno}}
\def\makeNormalizedRomappno{%
\expandafter\makeNormal\expandafter\normalizedromappno%
\expandafter{\romannumeral\appno}%
\edef\normalizedromappno{\uppercase{\normalizedromappno}}}
\def\Appendix#1{\global\advance\appno by1\relax\global\meqno=1\global\secno=0
\bigbreak\bigskip
\centerline{\twelvepoint \bf Appendix %
\romappno. #1}%
\par\nobreak\medskip\nobreak}
\def\tagappendix#1{\makeNormalizedRomappno%
\warnIfChanged#1{\normalizedromappno}%
\xdef#1{\normalizedromappno}%
\ifWritingAuxFile\immediate\write\auxfile{\noexpand\xdef\noexpand#1{#1}}\fi%
}
\def\appendix{\Appendix}

\def\eqn#1{\makeNormalizedRomappno%
\ifnum\secno>0%
  \warnIfChanged#1{\the\secno.\the\meqno}%
  \eqno(\the\secno.\the\meqno)\xdef#1{\the\secno.\the\meqno}%
     \global\advance\meqno by1
\else\ifnum\appno>0%
  \warnIfChanged#1{\normalizedromappno.\the\meqno}%
  \eqno({\rm\romappno}.\the\meqno)%
      \xdef#1{\normalizedromappno.\the\meqno}%
     \global\advance\meqno by1
\else%
  \warnIfChanged#1{\the\meqno}%
  \eqno(\the\meqno)\xdef#1{\the\meqno}%
     \global\advance\meqno by1
\fi\fi%
\eqlabeL#1%
\ifWritingAuxFile\immediate\write\auxfile{\noexpand\xdef\noexpand#1{#1}}\fi%
}
\def\defeqn#1{\makeNormalizedRomappno%
\ifnum\secno>0%
  \warnIfChanged#1{\the\secno.\the\meqno}%
  \xdef#1{\the\secno.\the\meqno}%
     \global\advance\meqno by1
\else\ifnum\appno>0%
  \warnIfChanged#1{\normalizedromappno.\the\meqno}%
  \xdef#1{\normalizedromappno.\the\meqno}%
     \global\advance\meqno by1
\else%
  \warnIfChanged#1{\the\meqno}%
  \xdef#1{\the\meqno}%
     \global\advance\meqno by1
\fi\fi%
\eqlabeL#1%
\ifWritingAuxFile\immediate\write\auxfile{\noexpand\xdef\noexpand#1{#1}}\fi%
}
\def\anoneqn{\makeNormalizedRomappno%
\ifnum\secno>0
  \eqno(\the\secno.\the\meqno)%
     \global\advance\meqno by1
\else\ifnum\appno>0
  \eqno({\rm\normalizedromappno}.\the\meqno)%
     \global\advance\meqno by1
\else
  \eqno(\the\meqno)%
     \global\advance\meqno by1
\fi\fi%
}
\def\mfig#1#2{\global\advance\figno by1%
\relax#1\the\figno%
\warnIfChanged#2{\the\figno}%
\edef#2{\the\figno}%
\reflabeL#2%
\ifWritingAuxFile\immediate\write\auxfile{\noexpand\xdef\noexpand#2{#2}}\fi%
}

\catcode`@=11 

\font\ninerm=cmr9
\font\eightrm=cmr8
\font\sixrm=cmr6

\def\loadtrueseventeenpoint{
 \font\seventeenrm=cmr10 at 17.28truept
 \font\seventeeni=cmmi10 at 17.28truept
 \font\seventeenbf=cmbx10 at 17.28truept
 \font\seventeenit=cmti10 at 17.28truept
 \font\seventeensl=cmsl10 at 17.28truept
 \font\seventeensy=cmsy10 at 17.28truept
}
\def\loadfourteenpoint{
\font\fourteenrm=cmr10 at 14.4pt
\font\fourteeni=cmmi10 at 14.4pt
\font\fourteenit=cmti10 at 14.4pt
\font\fourteensl=cmsl10 at 14.4pt
\font\fourteensy=cmsy10 at 14.4pt
\font\fourteenbf=cmbx10 at 14.4pt
}
\def\loadtruetwelvepoint{
\font\twelverm=cmr10 at 12truept
\font\twelvei=cmmi10 at 12truept
\font\twelveit=cmti10 at 12truept
\font\twelvesl=cmsl10 at 12truept
\font\twelvesy=cmsy10 at 12truept
\font\twelvebf=cmbx10 at 12truept
}

\font\ninei=cmmi9
\font\eighti=cmmi8
\font\sixi=cmmi6
\skewchar\ninei='177 \skewchar\eighti='177 \skewchar\sixi='177

\font\ninesy=cmsy9
\font\eightsy=cmsy8
\font\sixsy=cmsy6
\skewchar\ninesy='60 \skewchar\eightsy='60 \skewchar\sixsy='60

\font\ninebf=cmbx9
\font\eightbf=cmbx8
\font\sixbf=cmbx6

\font\ninett=cmtt9
\font\eighttt=cmtt8

\hyphenchar\tentt=-1 
\hyphenchar\ninett=-1
\hyphenchar\eighttt=-1

\font\ninesl=cmsl9
\font\eightsl=cmsl8

\font\nineit=cmti9
\font\eightit=cmti8


\newskip\ttglue
\def\tenpoint{\def\rm{\fam0\tenrm}%
  \textfont0=\tenrm \scriptfont0=\sevenrm \scriptscriptfont0=\fiverm
  \textfont1=\teni \scriptfont1=\seveni \scriptscriptfont1=\fivei
  \textfont2=\tensy \scriptfont2=\sevensy \scriptscriptfont2=\fivesy
  \textfont3=\tenex \scriptfont3=\tenex \scriptscriptfont3=\tenex
  \def\it{\fam\itfam\tenit}\textfont\itfam=\tenit
  \def\sl{\fam\slfam\tensl}\textfont\slfam=\tensl
  \def\bf{\fam\bffam\tenbf}\textfont\bffam=\tenbf \scriptfont\bffam=\sevenbf
  \scriptscriptfont\bffam=\fivebf
  \normalbaselineskip=12pt
  \let\sc=\eightrm
  \let\big=\tenbig
  \setbox\strutbox=\hbox{\vrule height8.5pt depth3.5pt width\z@}%
  \normalbaselines\rm}

\def\twelvepoint{\def\rm{\fam0\twelverm}%
  \textfont0=\twelverm \scriptfont0=\ninerm \scriptscriptfont0=\sevenrm
  \textfont1=\twelvei \scriptfont1=\ninei \scriptscriptfont1=\seveni
  \textfont2=\twelvesy \scriptfont2=\ninesy \scriptscriptfont2=\sevensy
  \textfont3=\tenex \scriptfont3=\tenex \scriptscriptfont3=\tenex
  \def\it{\fam\itfam\twelveit}\textfont\itfam=\twelveit
  \def\sl{\fam\slfam\twelvesl}\textfont\slfam=\twelvesl
  \def\bf{\fam\bffam\twelvebf}\textfont\bffam=\twelvebf
  \scriptfont\bffam=\ninebf
  \scriptscriptfont\bffam=\sevenbf
  \normalbaselineskip=12pt
  \let\sc=\eightrm
  \let\big=\tenbig
  \setbox\strutbox=\hbox{\vrule height8.5pt depth3.5pt width\z@}%
  \normalbaselines\rm}

\def\fourteenpoint{\def\rm{\fam0\fourteenrm}%
  \textfont0=\fourteenrm \scriptfont0=\tenrm \scriptscriptfont0=\sevenrm
  \textfont1=\fourteeni \scriptfont1=\teni \scriptscriptfont1=\seveni
  \textfont2=\fourteensy \scriptfont2=\tensy \scriptscriptfont2=\sevensy
  \textfont3=\tenex \scriptfont3=\tenex \scriptscriptfont3=\tenex
  \def\it{\fam\itfam\fourteenit}\textfont\itfam=\fourteenit
  \def\sl{\fam\slfam\fourteensl}\textfont\slfam=\fourteensl
  \def\bf{\fam\bffam\fourteenbf}\textfont\bffam=\fourteenbf%
  \scriptfont\bffam=\tenbf
  \scriptscriptfont\bffam=\sevenbf
  \normalbaselineskip=17pt
  \let\sc=\elevenrm
  \let\big=\tenbig
  \setbox\strutbox=\hbox{\vrule height8.5pt depth3.5pt width\z@}%
  \normalbaselines\rm}

\def\seventeenpoint{\def\rm{\fam0\seventeenrm}%
  \textfont0=\seventeenrm \scriptfont0=\fourteenrm \scriptscriptfont0=\tenrm
  \textfont1=\seventeeni \scriptfont1=\fourteeni \scriptscriptfont1=\teni
  \textfont2=\seventeensy \scriptfont2=\fourteensy \scriptscriptfont2=\tensy
  \textfont3=\tenex \scriptfont3=\tenex \scriptscriptfont3=\tenex
  \def\it{\fam\itfam\seventeenit}\textfont\itfam=\seventeenit
  \def\sl{\fam\slfam\seventeensl}\textfont\slfam=\seventeensl
  \def\bf{\fam\bffam\seventeenbf}\textfont\bffam=\seventeenbf%
  \scriptfont\bffam=\fourteenbf
  \scriptscriptfont\bffam=\twelvebf
  \normalbaselineskip=21pt
  \let\sc=\fourteenrm
  \let\big=\tenbig
  \setbox\strutbox=\hbox{\vrule height 12pt depth 6pt width\z@}%
  \normalbaselines\rm}

\def\ninepoint{\def\rm{\fam0\ninerm}%
  \textfont0=\ninerm \scriptfont0=\sixrm \scriptscriptfont0=\fiverm
  \textfont1=\ninei \scriptfont1=\sixi \scriptscriptfont1=\fivei
  \textfont2=\ninesy \scriptfont2=\sixsy \scriptscriptfont2=\fivesy
  \textfont3=\tenex \scriptfont3=\tenex \scriptscriptfont3=\tenex
  \def\it{\fam\itfam\nineit}\textfont\itfam=\nineit
  \def\sl{\fam\slfam\ninesl}\textfont\slfam=\ninesl
  \def\bf{\fam\bffam\ninebf}\textfont\bffam=\ninebf \scriptfont\bffam=\sixbf
  \scriptscriptfont\bffam=\fivebf
  \normalbaselineskip=11pt
  \let\sc=\sevenrm
  \let\big=\ninebig
  \setbox\strutbox=\hbox{\vrule height8pt depth3pt width\z@}%
  \normalbaselines\rm}

\def\eightpoint{\def\rm{\fam0\eightrm}%
  \textfont0=\eightrm \scriptfont0=\sixrm \scriptscriptfont0=\fiverm%
  \textfont1=\eighti \scriptfont1=\sixi \scriptscriptfont1=\fivei%
  \textfont2=\eightsy \scriptfont2=\sixsy \scriptscriptfont2=\fivesy%
  \textfont3=\tenex \scriptfont3=\tenex \scriptscriptfont3=\tenex%
  \def\it{\fam\itfam\eightit}\textfont\itfam=\eightit%
  \def\sl{\fam\slfam\eightsl}\textfont\slfam=\eightsl%
  \def\bf{\fam\bffam\eightbf}\textfont\bffam=\eightbf \scriptfont\bffam=\sixbf%
  \scriptscriptfont\bffam=\fivebf%
  \normalbaselineskip=9pt%
  \let\sc=\sixrm%
  \let\big=\eightbig%
  \setbox\strutbox=\hbox{\vrule height7pt depth2pt width\z@}%
  \normalbaselines\rm}

\def\tenbig#1{{\hbox{$\left#1\vbox to8.5pt{}\right.\n@space$}}}
\def\ninebig#1{{\hbox{$\textfont0=\tenrm\textfont2=\tensy
  \left#1\vbox to7.25pt{}\right.\n@space$}}}
\def\eightbig#1{{\hbox{$\textfont0=\ninerm\textfont2=\ninesy
  \left#1\vbox to6.5pt{}\right.\n@space$}}}

\def\footnote#1{\edef\@sf{\spacefactor\the\spacefactor}#1\@sf
      \insert\footins\bgroup\eightpoint
      \interlinepenalty100 \let\par=\endgraf
        \leftskip=\z@skip \rightskip=\z@skip
        \splittopskip=10pt plus 1pt minus 1pt \floatingpenalty=20000
        \smallskip\item{#1}\bgroup\strut\aftergroup\@foot\let\next}
\skip\footins=12pt plus 2pt minus 4pt 
\dimen\footins=30pc 

\newinsert\margin
\dimen\margin=\maxdimen
\def\titlefont{\seventeenpoint}
\loadtruetwelvepoint 
\loadtrueseventeenpoint
\catcode`\@=\active
\catcode`@=12  
\catcode`\"=\active

\def\eatOne#1{}
\def\ifundef#1{\expandafter\ifx%
\csname\expandafter\eatOne\string#1\endcsname\relax}
\def\notTrue{\iffalse}\def\isTrue{\iftrue}
\def\ifdef#1{{\ifundef#1%
\aftergroup\notTrue\else\aftergroup\isTrue\fi}}
\def\use#1{\ifundef#1\linemessage{Warning: \string#1 is undefined.}%
{\tt \string#1}\else#1\fi}


\global\newcount\refno \global\refno=1
\newwrite\rfile
\newlinechar=`\^^J
\def\ref#1#2{\the\refno\nref#1{#2}}
\def\nref#1#2{\xdef#1{\the\refno}%
\ifnum\refno=1\immediate\openout\rfile=refs.tmp\fi%
\immediate\write\rfile{\noexpand\item{[\noexpand#1]\ }#2.}%
\global\advance\refno by1}
\def\lref#1#2{\the\refno\xdef#1{\the\refno}%
\ifnum\refno=1\immediate\openout\rfile=refs.tmp\fi%
\immediate\write\rfile{\noexpand\item{[\noexpand#1]\ }#2\semi}%
\global\advance\refno by1}
\def\cref#1{\immediate\write\rfile{#1\semi}}
\def\eref#1{\immediate\write\rfile{#1.}}

\def\semi{;\hfil\noexpand\break}

\def\inputAuxIfPresent#1{\immediate\openin1=#1
\ifeof1\message{No file \auxfileName; I'll create one.
}\else\closein1\relax\input\auxfileName\fi%
}
\def\NPB{Nucl.\ Phys.\ B}
\def\PRL{Phys.\ Rev.\ Lett.\ }
\def\PRD{Phys.\ Rev.\ D}

\newif\ifWritingAuxFile
\newwrite\auxfile
\def\SetUpAuxFile{%
\xdef\auxfileName{\jobname.aux}%
\inputAuxIfPresent{\auxfileName}%
\WritingAuxFiletrue%
\immediate\openout\auxfile=\auxfileName}

\def\L{\left(}\def\R{\right)}
\def\LP{\left.}\def\RP{\right.}
\def\LB{\left[}\def\RB{\right]}

\def\bye{\par\vfill\supereject%
\ifAnyCounterChanged\linemessage{
Some counters have changed.  Re-run tex to fix them up.}\fi%
\end}
%
\def\L{\left(}
\def\R{\right)}

\def\Tr{\mathop{\rm Tr}\nolimits}
\def\Gr{\mathop{\rm Gr}\nolimits}

\def\Re{\mathop{\rm Re}\nolimits}

\def\si{\sigma}

\def\eps{\epsilon}

\def\LP{\left.}\def\RP{\right.}

\def\dl^#1_#2{\delta^{#1}{}_{#2}}

\def\Li{\mathop{\rm Li}\nolimits}

\def\A#1{{\cal A}_{#1}}

\def\si{\sigma}

\def\Tr{\mathop{\rm Tr}\nolimits}

\def\A#1{{\cal A}_{#1}}

\def\Re{\mathop{\rm Re}}
\def\L{\left(}\def\R{\right)}
\def\LP{\left.}\def\RP{\right.}
\def\spa#1.#2{\left\langle#1\,#2\right\rangle}
\def\spb#1.#2{\left[#1\,#2\right]}
\def\lor#1.#2{\left(#1\,#2\right)}
\def\sand#1.#2.#3{%
\left\langle\smash{#1}{\vphantom1}^{-}\right|{#2}%
\left|\smash{#3}{\vphantom1}^{-}\right\rangle}
\def\sandp#1.#2.#3{%
\left\langle\smash{#1}{\vphantom1}^{-}\right|{#2}%
\left|\smash{#3}{\vphantom1}^{+}\right\rangle}
\def\sandpp#1.#2.#3{%
\left\langle\smash{#1}{\vphantom1}^{+}\right|{#2}%
\left|\smash{#3}{\vphantom1}^{+}\right\rangle}

\SetUpAuxFile
\loadfourteenpoint
\nopagenumbers\hsize=\hstitle\vskip1in
\overfullrule 0pt
\hfuzz 35 pt
\vbadness=10001
%
%

\def\Tr{\mathop{\rm Tr}\nolimits}

\def\Li{\mathop{\rm Li}\nolimits}

\def\e{\epsilon}

\def\"#1{{\accent127 #1}}

\def\lr{\leftrightarrow}
\def\spa#1.#2{\left\langle\mskip-1mu#1\,#2\mskip-1mu\right\rangle}
\def\spb#1.#2{\left[\mskip-1mu#1\,#2\mskip-1mu\right]}

\def\cg{c_\Gamma}

\def\Ls{\mathop{\rm Ls}\nolimits}
\def\Ll{\mathop{\rm L}\nolimits}
\def\Null{\mathop{}\nolimits}
\def\lv{\varepsilon}


\rightline{CERN-TH.6803/93}
\rightline{SLAC--PUB--6054}
\rightline{UCLA/93/TEP/4}
\rightline{February, 1993}
\rightline{  }

\leftlabelstrue
\vskip -1.0 in
\Title{One-Loop Corrections to Five-Gluon Amplitudes}

\centerline{Zvi Bern${}^{\ddagger}$}
\baselineskip12truept
\centerline{\it Department of Physics}
\centerline{\it University of California, Los Angeles}
\centerline{\it Los Angeles, CA 90024}
\centerline{\tt bern@physics.ucla.edu}

\smallskip\smallskip

\baselineskip17truept
\centerline{Lance Dixon${}^{\star}$}
\baselineskip12truept
\centerline{\it Stanford Linear Accelerator Center}
\centerline{\it Stanford, CA 94309}
\centerline{\tt lance@slacvm.slac.stanford.edu}

\smallskip \centerline{and} \smallskip

\baselineskip17truept
\centerline{David A. Kosower${}^{\dagger}$}
\baselineskip12truept
\centerline{\it Theory Division}
\centerline{\it CERN}
\centerline{\it CH-1211 Geneva 23}
\centerline{\it Switzerland}
\centerline{\tt kosower@nxth02.cern.ch}

\vskip 0.2in\baselineskip13truept

\vskip 0.75truein
\centerline{\bf Abstract}

{\narrower

We present the one-loop helicity amplitudes with five external gluons.
The computation employs string-based methods, new techniques for
performing tensor integrals, and improvements in
the spinor helicity method.
}

\vskip 0.3truein

\centerline{\sl Submitted to Physical Review Letters}

\vfill
\vskip 0.1in
\noindent\hrule width 3.6in\hfil\break
${}^{\ddagger}$Research supported by the Texas National Research
Laboratory Commission grant FCFY9202.\hfil\break
${}^{\star}$Research supported by the Department of
Energy under grant DE-AC03-76SF00515.\hfil\break
${}^{\dagger}$On leave from the Centre d'Etudes of Saclay,
F-91191 Gif-sur-Yvette cedex, France.\hfil\break
%
\Date{}

\line{}


Calculations beyond the leading order in quantum chromodynamics
are important to refining our understanding of known physics in
present-day and future collider experiments, such as the
Tevatron or the SSC and LHC.  In jet physics,
next-to-leading order
calculations are important to curing several deficiencies of
their leading-order counterparts: the strong spurious
dependence on the renormalization scale;
the lack of sensitivity to the jet resolution parameters,
namely the minimum transverse energy and
the jet cone size; and the absence of a warning about
dangerous infrared logarithms.
The one-loop corrections
to matrix elements for $2\rightarrow 2$ processes in QCD,
a key ingredient of the next-to-leading order calculations
of inclusive-jet and two-jet cross-sections and distributions,
were computed by Ellis and Sexton~[\ref\ES{R. K. Ellis and J. C. Sexton,
Nucl.\ Phys.\ B269:445 (1986)}].
To go beyond these cross-sections,
whether to higher orders for two-jet cross-sections,
or to next-to-leading order for three-jet cross-sections and
distributions, requires the computation of the one-loop
corrections to the $2\rightarrow 3$ matrix elements.
At hadron colliders, the QCD coupling $\alpha_s$,
and the manner of its running, can be extracted from purely
hadronic processes by comparing three-jet production to
two-jet production, at various center-of-mass energies.
The presence of infrared logarithms in both of these quantities
means that this cannot be done sensibly unless both quantities
are known to next-to-leading order.

We present here the one-loop matrix elements with five external gluons,
which are the hardest part of a $2\rightarrow 3$ calculation
if a traditional diagrammatic method is used.
We have performed the calculation using the string-based
methods developed in
ref.~[\ref\StringMethod{%
Z. Bern and D. A. Kosower, \PRL 66:1669 (1991); \NPB 379:451 (1992)}]
as more efficient tools for one-loop calculations with external
gluons.
The rules presented there were derived by taking the infinite-tension
limit of an appropriately-constructed heterotic string amplitude.
The structure of the rules can also be understood in conventional
field theory~[\ref\FieldTheory{Z. Bern and D. C. Dunbar,
 \NPB379:562 (1992)}], and
the application of such methods to a calculation such as the present
one does not require any knowledge of string theory.
(It turns out that it is possible to construct a set of rules
yielding more compact integral representation of gluon amplitudes
at intermediate stages than would emerge from a straightforward
application of the rules in ref.~[\use\StringMethod].  This alternate
set of string-based rules will be discussed elsewhere.)

In the string-based method, one first decomposes the $n$-gluon amplitude,
depending on the external momenta, helicities, and color
indices $k_i$, $\lambda_i$, and $a_i$,
into sums over certain permutations of color factors, times partial
amplitudes, in analogy to the
helicity~[\lref\SpinorHelicity{%
   F.\ A.\ Berends, R.\ Kleiss, P.\ De Causmaecker, R.\ Gastmans,
   and T.\ T.\ Wu, Phys.\ Lett.\ 103B:124 (1981)\semi
   P.\ De Causmaeker, R.\ Gastmans,  W.\ Troost,
   and  T.\ T.\ Wu, Nucl. Phys. B206:53 (1982)}%
  \eref{R.\ Kleiss and W.\ J.\ Stirling,
   Nucl.\ Phys.\ B262:235 (1985)\semi
   J.\ F.\ Gunion and Z.\ Kunszt, Phys.\ Lett.\ 161B:333 (1985)\semi
   R.\ Gastmans and T.T.\ Wu,
   {\it The Ubiquitous Photon: Helicity Method for QED and QCD}
   (Clarendon Press) (1990)\semi
   M. Mangano and S. J. Parke, Phys. Rept. 200:301 (1991)},%
\ref\XZC{Z. Xu, D.-H.\ Zhang, L. Chang, Tsinghua University
                  preprint TUTP--84/3 (1984), unpublished\semi
Z.\ Xu, D. Zhang, L. Chang, Nucl.\ Phys.\ B291:392 (1987)}]
and
color~[\ref\TreeColor{F. A. Berends and W. T. Giele, Nucl. Phys.
     B294:700 (1987)\semi
     M. Mangano, S. J. Parke and Z. Xu, Nucl.\ Phys.\ B298:653 (1988);
     Nucl.\ Phys.\ B299:673 (1988)\semi
     D. Zeppenfeld, Int. J. Mod. Phys. A3:2175 (1988)}]
decomposition of tree amplitudes.
At one-loop order in an $SU(N)$ theory,
one must also sum over the different spins $J$ of the internal particles;
this takes the following form when all internal particles transform
as color adjoints,
$$
{\cal A}_n\L \{k_,\lambda_i,a_i\}\R =
  \sum_{J} n_J\,\sum_{c=1}^{\lfloor{n/2}\rfloor+1}
      \sum_{\sigma \in S_n/S_{n;c}}
     \Gr_{n;c}\L \sigma \R\,A_{n;c}^{[J]}(\sigma)
\eqn\ColorDecomposition$$
where $\Gr_{n;1}(1) = N \Tr\L T^{a_1}\cdots T^{a_n}\R$,
$\Gr_{n;c}(1) = \Tr\L T^{a_1}\cdots T^{a_{c-1}}\R\,
\Tr\L T^{a_c}\cdots T^{a_n}\R$, $S_n$ is the set of all permutations of
$n$ objects, and $S_{n;c}$ is the subset leaving the trace structure
$\Gr_{n;c}$ invariant.  The $T^a$ are the set of hermitian traceless
$N\times N$ matrices, normalized so that $\Tr\L T^a T^b\R = \delta^{ab}$.
For internal particles in the fundamental ($N+\bar{N}$) representation,
only the single-trace color structure ($c=1$) is present,
and it is smaller by a factor of $N$.
We take in each case a spin-$J$ particle with two states: gauge bosons,
Weyl fermions, and complex scalars.

The objects one calculates are the partial amplitudes
$A_{n;c}^{[J]}$, which depend
only on the external momenta and helicities.
For the five-point
function, there is only one independent partial amplitude for each
configuration of external helicities; $A_{5;2}$ and $A_{5;3}$
are related to the adjoint contributions to $A_{5;1}$ via decoupling
equations~[\ref\Color{Z. Bern and D. A. Kosower, \NPB 362:389 (1991)}].

The string-based method meshes well with
the spinor helicity representation for
the polarization vectors~[\SpinorHelicity,\XZC], which provides
an efficient method for extracting the essential gauge-invariant
information in an on-shell amplitude.  This method yields expressions
written in terms of spinor products $\spa{i}.{j}$ and $\spb{i}.{j}$,
which are square roots of
Lorentz products $s_{ij} = (k_i+k_j)^2$ (up to a phase).
Unfortunately, the relations
--- momentum conservation and the Schouten identity ---
between different forms of a given expression are nonlinear, which
makes it hard to give a canonical form for such expressions, or
equivalently makes it hard to simplify complicated expressions.
However, one can evaluate ``phase-invariant'' combinations of spinor
products in terms of $s_{ij}$ and contractions of the Levi-Civita
tensor $\lv(i,j,m,n) = 4i\varepsilon_{\mu\nu\rho\sigma}
        k_i^\mu k_j^\nu k_m^\rho k_n^\sigma
    \ =\ \spb{i}.{j}\spa{j}.{m}\spb{m}.{n}\spa{n}.{i}
       - \spa{i}.{j}\spb{j}.{m}\spa{m}.{n}\spb{n}.{i}$.
It suffices to calculate the ratios
$$
  {\spa{i}.{j}\spb{j}.{k}\over\spa{i}.{l}\spb{l}.{k}}\ =\
 {s_{il}s_{jk}+s_{kl}s_{ij}-s_{ik}s_{jl}-\lv(i,j,k,l)
    \over 2s_{il}s_{kl}}\ ,
\eqn\spinorsimp
$$
using e.g. methods in
reference~[\ref\Fearing{H. W. Fearing and R. R. Silbar,
Phys. Rev. D6:471 (1971)}].
In this way spinor products can be eliminated from any
expression, apart from an overall prefactor.

For massless five-point kinematics, such an expression can then be
written as a rational function in the five kinematic variables
$\{\beta_1,\beta_2^*,\beta_3,\beta_4^*,\beta_5\}$
(or any cyclic permutation of this set), where
$$
  \beta_i\ =\ \spb{i,}.{i+1}\spa{i+1,}.{i+2}\spb{i+2,}.{i+3}
 \spa{i+3,}.{i}
  \times \Bigl(-\prod_{j=1}^5 s_{j,j+1}\Bigr)^{-1/2}\ .
\eqn\betadefn
$$
The only independent Levi-Civita contraction is given by
$\lv(1,2,3,4)/(-\prod_{j=1}^5 s_{j,j+1})^{1/2}
 = (\beta_5\beta_2^* + \beta_1\beta_3 + \beta_2^*\beta_4^*
  + \beta_3\beta_5 + \beta_4^*\beta_1)/\beta_3
 = \beta_i-\beta_i^*$ for any $i$,
and the independent Lorentz products by
$s_{i,i+1} = -1/((\beta_i+\beta_{i+1}^*)(\beta_{i+2}+\beta_{i+3}^*))$.
Simplification of rational functions in $\beta_i$ is straightforward.

The $\beta_i$ variables are related to the variables
$\gamma_i$ and $\hat\Delta_5$ used in
ref.~[\ref\Integrals{Z. Bern, L. Dixon, and D. A. Kosower,
preprints SLAC--PUB--6001 and SLAC--PUB--5947}]
to perform pentagon integrals, via
$\beta_i^{(*)} = -(\gamma_{i+2}\pm\sqrt{\hat\Delta_5})/2$.
Indeed, the derivative approach to evaluating tensor
integrals~[\Integrals], when applied to the pentagon integrands
encountered in the five-gluon calculation,
and expressed in terms of the appropriate set of
$\beta_i$ variables, allows one to significantly
reduce the degree and size of the Feynman parameter polynomials
in the integrand.

At tree-level, certain helicity amplitudes vanish
identically~[\ref\ParkeTaylor{S. J. Parke and T. R. Taylor,
Phys.\ Rev.\ Lett.\ 56:2459 (1986)}].  The corresponding
one-loop amplitudes are then free of infrared divergences.
The remaining amplitudes are infrared-divergent;
for practical purposes these divergences
must be regulated using dimensional regularization.  The computation
of these helicity amplitudes thus requires the knowledge of
five-point loop integrals in $D=4-2\e$~[\Integrals,
  \ref\EGY{R. K. Ellis, W. T. Giele and E. Yehudai, in preparation}].

\def\v{V}
\def\f{F}
For the finite helicity amplitudes, supersymmetric
identities~[\ref\SUSYid{M. T. Grisaru and H. N. Pendleton,
  Nucl. Phys. B124:81 (1977)}]
imply that the contributions
of particles of different spin circulating around the loop are related,
$A_{n;c}^{[1]} = -A_{n;c}^{[1/2]} = A_{n;c}^{[0]}$.
(This holds true for the partial amplitudes whether or not the
theory as a whole is supersymmetric.)  Indeed, in the string-based
method, these identities hold for the integrands of each diagram.
The amplitudes are
$$\eqalign{
A_{5;1}^{[1]}\L 1^{+},2^{+},3^{+},4^{+},5^{+}\R &=
{i\over96\pi^2}
  {s_{12}s_{23}+s_{23}s_{34}+s_{34}s_{45}+s_{45}s_{51}+s_{51}s_{12}
   +\lv(1,2,3,4)\over
   \spa1.2\spa2.3\spa3.4\spa4.5\spa5.1}\,,\cr
A_{5;1}^{[1]}\L 1^{-},2^{+},3^{+},4^{+},5^{+}\R &=
{i\over48\pi^2}
   {1\over\spb1.2\spa2.3\spa3.4\spa4.5\spb5.1} \LB\vphantom{\sum}
      (s_{23}+s_{34}+s_{45}) {\spb2.5}^2
  - \spb2.4\spa4.3\spb3.5\spb2.5
       \vphantom{\L
         {\spa1.2}^2 {\spa1.3}^2 {\spb2.3\over\spa2.3}\R}\RP\cr
  &\qquad \LP\vphantom{\sum}
   - {\spb1.2\spb1.5\over\spa1.2\spa1.5}
       \L
         {\spa1.2}^2 {\spa1.3}^2 {\spb2.3\over\spa2.3}
        +{\spa1.3}^2 {\spa1.4}^2 {\spb3.4\over\spa3.4}
        +{\spa1.4}^2 {\spa1.5}^2 {\spb4.5\over\spa4.5}\R
    \RB\,.\cr
}\anoneqn$$

\def\Atree{A^{\rm tree}}
In order to present the results for the remaining, infrared-divergent
amplitudes in a compact form, it is helpful to
define the following functions,
$$\eqalign{
\Ll_0(r) &= {\ln(r)\over 1-r}\,,\hskip 10mm
\Ll_1(r) = {\ln(r)+1-r\over (1-r)^2}\,,\hskip 10mm
\Ll_2(r) = {\ln(r)-(r-1/r)/2\over (1-r)^3}\,,\cr
\Ls_1(r_1,r_2) &= {1\over (1-r_1-r_2)^2}
\LB \Li_2(1-r_1) + \Li_2(1-r_2) + \ln r_1\,\ln r_2 - {\pi^2\over6}\RP\cr
&\hskip 15mm\LP\vphantom{{\pi^2\over6}}
   + (1-r_1-r_2) \L \Ll_0(r_1) + \Ll_0(r_2)\R
\RB\ ,\cr
}\eqn\Lsdef$$%
where $\Li_2$ is the dilogarithm; a prefactor,
$$\eqalign{
  c_\Gamma &= {(4\pi)^\eps\over16\pi^2}
              {\Gamma(1+\eps)\Gamma^2(1-\eps)\over\Gamma(1-2\eps)}\ ,
    \cr
}\anoneqn$$
a universal function,
$$\eqalign{
\v^{g} &= -{1\over\e^2}\sum_{j=1}^5 \L {\mu^2\over -s_{j,j+1}}\R^\e
          +\sum_{j=1}^5 \ln\L{-s_{j,j+1}\over -s_{j+1,j+2}}\R\,
                        \ln\L{-s_{j+2,j-2}\over -s_{j-2,j-1}}\R
          +{5\over6}\pi^2-{\delta_R\over3}\,,
  \cr
}\anoneqn$$
the following functions for the $(1^{-},2^{-},3^{+},4^{+},5^{+})$
helicity configuration,
$$\eqalign{
\v^{f}&= -{5\over2\e}-{1\over2}\LB\ln\L{\mu^2\over -s_{23}}\R
                                 +\ln\L{\mu^2\over -s_{51}}\R\RB-2,
\hskip 10mm
\v^{s} = -{1\over3} \v^{f} + {2\over9}\cr
\f^{f} &= -{1\over 2}
   {{\spa1.2}^2 \L\spa2.3\spb3.4\spa4.1+\spa2.4\spb4.5\spa5.1\R\over
    \spa2.3\spa3.4\spa4.5\spa5.1}
     {\Ll_0\L {-s_{23}\over -s_{51}}\R\over s_{51}}\cr
\f^{s} &=
    -{1\over 3}
   {\spb3.4\spa4.1\spa2.4\spb4.5
     \L\spa2.3\spb3.4\spa4.1+\spa2.4\spb4.5\spa5.1\R\over\spa3.4\spa4.5}
     {\Ll_2\L {-s_{23}\over -s_{51}}\R\over s_{51}^3}
     - {1\over3}\f^{f}\cr
 & \hskip 3mm
   -{1\over3}{\spa3.5{\spb3.5}^3\over\spb1.2\spb2.3\spa3.4\spa4.5\spb5.1}
     +{1\over3}{\spa1.2{\spb3.5}^2\over\spb2.3\spa3.4\spa4.5\spb5.1}
     +{1\over6}{\spa1.2\spb3.4\spa4.1\spa2.4\spb4.5\over
                  s_{23}\spa3.4\spa4.5 s_{51}}\,,\cr
}\anoneqn$$
and the corresponding ones for the $(1^{-},2^{+},3^{-},4^{+},5^{+})$
helicity configuration,
$$\hskip -10pt\eqalign{
\v^{f}&= -{5\over2\e}-{1\over2}\LB\ln\L{\mu^2\over -s_{34}}\R
                                 +\ln\L{\mu^2\over -s_{51}}\R\RB-2,
\hskip 10mm
\v^{s} = -{1\over3} \v^{f} +{2\over9}\cr
\f^{f} &=
      -{{\spa1.3}^2 {\spa4.1} {\spb2.4}^2
         \over {\spa4.5} {\spa5.1}}
           {\Ls_1\L {-s_{23}\over -s_{51}},\,{-s_{34}\over -s_{51}}\R
            \over s_{51}^2}
      +{{\spa1.3}^2 {\spa5.3} {\spb2.5}^2
         \over {\spa3.4} {\spa4.5}}
          {\Ls_1\L {-s_{12}\over -s_{34}},\,{-s_{51}\over -s_{34}}\R
           \over s_{34}^2}\cr
&\hskip 15mm
      -{1\over2} {{\spa1.3}^3
          (\spa1.5 \spb5.2 \spa2.3-\spa3.4 \spb4.2 \spa2.1)
             \over \spa1.2\spa2.3\spa3.4\spa4.5\spa5.1}
           {\Ll_0\L {-s_{34}\over -s_{51}}\R\over s_{51}}\cr
\f^{s} &=
      - {{\spa1.2} {\spa2.3} {\spa3.4}
          {\spa4.1}^2 {\spb2.4}^2
         \over {\spa4.5} {\spa5.1} {\spa2.4}^2}\,
          {2 \, \Ls_1\L {-s_{23}\over -s_{51}},\,{-s_{34}\over -s_{51}}\R
           + \Ll_1\L {-s_{23}\over -s_{51}}\R
           + \Ll_1\L {-s_{34}\over -s_{51}}\R  \over s_{51}^2} \cr
& \hskip 4mm
      + {{\spa3.2} {\spa2.1} {\spa1.5}
          {\spa5.3}^2 {\spb2.5}^2
         \over {\spa5.4} {\spa4.3} {\spa2.5}^2}\,
          {2 \, \Ls_1\L {-s_{12}\over -s_{34}},\,{-s_{51}\over -s_{34}}\R
           + \Ll_1\L {-s_{12}\over -s_{34}}\R
           + \Ll_1\L {-s_{51}\over -s_{34}}\R  \over s_{34}^2} \cr
& \hskip 4mm
      +{2\over 3} {{\spa2.3}^2 {\spa4.1}^3 {\spb2.4}^3
          \over {\spa4.5} {\spa5.1} {\spa2.4}}
          {\Ll_2\L {-s_{23}\over -s_{51}}\R  \over s_{51}^3}
      -{2\over 3} {{\spa2.1}^2 {\spa5.3}^3 {\spb2.5}^3
          \over {\spa5.4} {\spa4.3} {\spa2.5}}
          {\Ll_2\L {-s_{12}\over -s_{34}}\R  \over s_{34}^3} \cr
& \hskip 4mm + {\Ll_2\L {-s_{34}\over -s_{51}}\R\over s_{51}^3}\,
     \biggl( {1\over3} { \spa1.3\spb2.4\spb2.5
     (\spa1.5 \spb5.2 \spa2.3-\spa3.4 \spb4.2 \spa2.1) \over \spa4.5}
  \cr
&\hskip 4mm
    +{2\over 3} {{\spa1.2}^2{\spa3.4}^2\spa4.1{\spb2.4}^3
            \over \spa4.5\spa5.1\spa2.4}
    -{2\over 3} {{\spa3.2}^2{\spa1.5}^2\spa5.3{\spb2.5}^3
            \over \spa5.4\spa4.3\spa2.5}  \biggr) \cr
&\hskip 4mm
   +{1\over 6} {{\spa1.3}^3
        \L \spa1.5\spb5.2\spa2.3 - \spa3.4\spb4.2\spa2.1\R
          \over \spa1.2\spa2.3\spa3.4\spa4.5\spa5.1}\,
            {\Ll_0\L {-s_{34}\over -s_{51}}\R\over s_{51}}
   +{1\over 3} {{\spb2.4}^2 {\spb2.5}^2
          \over {\spb1.2} {\spb2.3} {\spb3.4} {\spa4.5} {\spb5.1}}
            \cr
&\hskip 4mm
   -{1\over 3} {{\spa1.2} {\spa4.1}^2 {\spb2.4}^3
          \over {\spa4.5} {\spa5.1} {\spa2.4} {\spb2.3} {\spb3.4} s_{51}}
   +{1\over 3} {{\spa3.2} {\spa5.3}^2 {\spb2.5}^3
          \over {\spa5.4} {\spa4.3} {\spa2.5} {\spb2.1} {\spb1.5} s_{34}}
   +{1\over 6}
    {{\spa1.3}^2 \spb2.4\spb2.5 \over s_{34} \spa4.5 s_{51}}
           \ . \cr
}\anoneqn$$
For positive values of $s_{ij}$, the logarithms and dilogarithms
develop imaginary parts according to the prescription
$s_{ij} \to s_{ij} + i\varepsilon$.
We also remind the reader of the tree amplitudes,
$\Atree_5\L 1^{-},2^{-},3^{+},4^{+},5^{+}\R
  = i{\spa1.2}^4/(\spa1.2\spa2.3\spa3.4\spa4.5\spa5.1)$ and
$\Atree_5\L 1^{-},2^{+},3^{-},4^{+},5^{+}\R
  = i{\spa1.3}^4/(\spa1.2\spa2.3\spa3.4\spa4.5\spa5.1)$.

In terms of these functions, the $\overline{{\rm MS}}$
renormalized amplitudes are
$$\eqalign{
A_{5;1}^{[0]} &= \cg \L \v^s \Atree_5 + i \f^s \R\,,\cr
A_{5;1}^{[1/2]} &= -\cg \L (\v^f+\v^s) \Atree_5 + i(\f^f+\f^s) \R\,,\cr
A_{5;1}^{[1]} &= \cg \L (\v^g+4\v^f+\v^s) \Atree_5 + i (4\f^f+\f^s) \R
  \,.\cr
}\eqn\Totalamp$$
The rest of the helicity amplitudes are related by cyclic permutations
or complex conjugation to those given above.
It is interesting to note that in supersymmetric theories, the
$\v^s$ and $\f^s$ terms cancel out of the final amplitude, and that
in $N=4$ supersymmetric theories only the $\v^g$ term survives.
The separation implied above into $g$, $f$, and $s$ pieces arises
naturally on a diagram-by-diagram basis within the string-based
approach.  In this approach the $\v^g$ term represents the only
calculational difference between the contributions with
gluons circulating around the loop, and those with fermions;
this term has a particularly simple expression at
every intermediate stage of the calculation.
The parameter $\delta_R$ controls the variant of dimensional
regularization scheme~[\use\StringMethod]: for $\delta_R=0$, one
obtains the four-dimensional helicity
scheme, while for $\delta_R=1$ one obtains the 't~Hooft--Veltman
scheme.

There are several checks we have applied.  We have checked gauge
invariance, both by computing amplitudes with longitudinal
gluons, verifying that one obtains zero, and by calculating a
helicity amplitude with an alternate choice of spinor-helicity
reference momenta, and verifying that the result is unchanged.
In addition, the forms given above display manifestly the
reflection symmetries
expected of the amplitudes, symmetries that are not present in the
contributions of the individual diagrams.
The amplitudes also have consistent limits as one of the gluon
momenta becomes soft, and as two adjacent momenta become collinear.

\def\Atrees{A^{{\rm tree}\,*}}
\def\si{\sigma}\def\phstar{{\phantom{*}}}
\def\NLO{\rm NLO}
\def\Part#1#2{P\hbox{${#1\choose #2}$}}
\def\fiveperm#1#2#3#4#5{(#1\,#2\,#3\,#4\,#5)}
At next-to-leading order, only the infrared-divergent helicity
amplitudes~(\use\Lsdef--\use\Totalamp)
enter into the construction of a program for physical quantities.
In order to construct such a program for three-jet
quantities, one must form the interference
of the tree amplitude with the loop amplitude;
this has the form~[\use\Color]
$$\eqalign{
\sum_{\rm colors} \LB\A{5}^*\A{5}^\phstar\RB_{\NLO}
&= 2 g^{8} N^4 \L N^2 -1\R \LB
 \Re\sum_{\si\in S_5/Z_5}
\Atrees_5(\si) A_{5;1}(\si)\RP\cr
&+ {1\over N^2} \Re\sum_{\rho\in S_5/Z_5}
\LB\vphantom{\sum}
  \Atrees_5(r\cdot\rho) A_{5;1}(\rho)
  -\Atrees_5(\rho) A_{5;1}(r\cdot\rho)\RB\cr
&\LP + {2\over N^2} \Re\sum_{h\in H_5}
\sum_{p\in \Part{5}{3}}
\Atrees_5(h\cdot p) A_{5;3}(p)\RB \; ,\cr
}\eqn\NLOCrossSection$$
where $r$ is the permutation $\fiveperm24135$;
$\Part{5}{3}$ is the ten-element
set of distinct partitions of five elements into
lists of length two and three; and
$H_5 = \{ \fiveperm12345, \fiveperm34125,\Null \fiveperm31245,\Null
       \fiveperm21345,\Null \fiveperm32145,\Null\break \fiveperm34215\}$.
For QCD with $n_f$ flavors of massless quarks,
one substitutes
$A_{5;1} \to A_{5;1}^{[1]} + {n_f\over N} A_{5;1}^{[1/2]}$ and
$A_{5;3} \to A_{5;3}^{[1]}$ into equation~(\use\NLOCrossSection).
One must then combine this virtual correction with the singular terms in
the $2\rightarrow4$ matrix elements arising from the integration over
soft and collinear phase space.  The Giele-Glover
formalism~[\ref\GG{W. T. Giele and E. W. N. Glover,
\PRD 46:1980 (1992)\semi
W. T. Giele, E. W. N. Glover, and D. A. Kosower,
preprint Fermilab--Pub--92/230--T}] makes use of the
color ordering in construction of universal functions representing
the results of the soft and collinear integrations, and is the
most convenient one for doing so.  We have used it to check that the
poles in $\eps$ do cancel as expected.

\vfill\eject\immediate\closeout\rfile
\centerline{{\bf References}}\bigskip\frenchspacing%
\input refs.tmp\vfill\eject\nonfrenchspacing

\bye